\documentclass[11pt,twoside]{article}

\usepackage{asp2006}
\usepackage{epsf}
\usepackage{psfig}
\usepackage{lscape}
\begin{document}
\title{An X-ray view of Pictor A radio lobes: a spatially resolved study}   
\author{G. Migliori$^{1,2,3}$, P. Grandi$^{2}$, G.C.G. Palumbo$^{1}$, G. Brunetti$^{4}$,\\ C. Stanghellini$^{4}$}   
\affil{$^{1}$Dipartimento di Astronomia, Universit\`a di Bologna,
via Ranzani 1, 40127 Bologna, Italy\\
$^{2}$Istituto di Astrofisica e Fisica Cosmica-Bologna, INAF,
 Via Gobetti 101, I-40129 Bologna,Italy\\
$^{3}$SISSA/ISAS, Via Beirut 2-4, I-34014, Trieste, Italy\\
$^{4}$Istituto di Radioastronomia-Bologna,INAF, Via Gobetti 101, 
I-40129 Bologna,Italy}    

\begin{abstract} 
A spatially resolved analysis of the lobes of the radio galaxy Pictor A has been performed for the first time starting from a 50 ksec XMM-Newton observation. Magnetic field, $B_{IC}$, particle density, particle to magnetic field energy density ratios have been measured. 
 Our study shows that $B_{IC}$ varies through the lobes. 
On the contrary, a rather uniform distribution of the particles is observed. In both the lobes, the equipartition magnetic field, $B_{eq}$, is bigger than the Inverse Compton value, $B_{IC}$, calculated from the radio to X-ray flux ratio.
\end{abstract}


\section{Introduction}   
 The X-ray observatories {\it Chandra} and {\it XMM-Newton} have made possibile focusing on the X-ray extended emission from the faintest X-ray components of a radio galaxy, the lobes. The first aim has been to define the origin of the X-ray emission: thermal, due to external gas, compressed by the expansion of the lobes \citep{kra00,rey01}, or non-thermal, due to Inverse Compton (IC) by relativistic electrons with $\gamma \sim 10^{3}$ on CMB photons \citep{harr79,mil80}.
 In the last case, we can use X-ray fluxes together with radio fluxes to obtain a direct estimate of the average magnetic field ($B_{IC}$) along the line of side and, independently, the number densities of the IC emitting particles in the lobes. This method has allowed to go beyond and, at the same time, to check the validity of the classical assumption of the minimum energy condition (equipartition) in the extended regions. Even though the subject is still highly debated \citep{hard02,iso02,gra03,cro05}, however it seems reasonable to conclude that the equipartition argument represents a viable zero-order reference guide.
Also the study of the variations of the magnetic field and particle densities turns out to be useful to acquire informations on the dynamical plasma evolution in the lobes. Even though this procedure has been applied only to a few radio galaxies because of observational limits, the results indicate an enhancement of the magnetic energy density towards the edge of the lobes \citep{tash01,iso02,iso05}.\\
\section{Pictor A}
Here we present a study of the lobes of the radio galaxy Pictor A, based on a new {\it XMM-Newton} observation of 50 ksec. The origin of the X-ray emission was exhaustively discussed in some previous papers \citep{gra03, hard05}, and in \citet{migl06} we reported  the preliminary  spectral analysis from this last 50 ksec observation, conducted on two datasets accumulated on the east and west lobe regions (respectively E and W regions hereafter). The extended X-ray emission is confirmed to be due to non-thermal processes, i.e. IC on CMB photons (for a detailed discussion see \citet{migl07}, Fig.~\ref{fig_1} and Tab.~\ref{tab_1} for a recap of the spectral analysis). 
The originality of the work on this new observation rather resides in the possibility of realize a spatially resolved study of the lobes, investigating the possible variations of the physical conditions inside the lobes themselves. 
For this reason we accumulated EPIC/MOS1 data sets from selected rectangular regions, three for each lobe, labeled as described in Fig.~\ref{fig_1}. Since the low number of counts, we obtained X-ray fluxes, $F_{0.5-2~keV}$, for each subregion, once fixed the X-ray spectral slope, $\Gamma$, referring to the value, $\Gamma=1.8$, found for the spectra of the E and W regions.
\section{Discussion} 
\begin{figure}[!hb]
\caption{{\it (On the left)-}{\it XMM-Newton}/MOS1 image (0.2-10 keV) of Pictor A observed on January 2005. The blue  and the red circles (respectively on the left and on the right respect to the nuclear region) represent the east (E) and west (W) extraction regions of the lobes used for the spectral analysis. The nucleus, two hot spots, jet contribution and a point source (black circles) are excluded. {\it (On the right)-}VLA map \citep{perl97} at 20 cm. Rectangular (pink) boxes represent the subregions used for the spatially resolved analysis: e1, e2, e3 for the east lobe, w1, w2, w3, for the west lobe.  Again, black circle delimitates the excluded nuclear region.}
\label{fig_1}
\plottwo{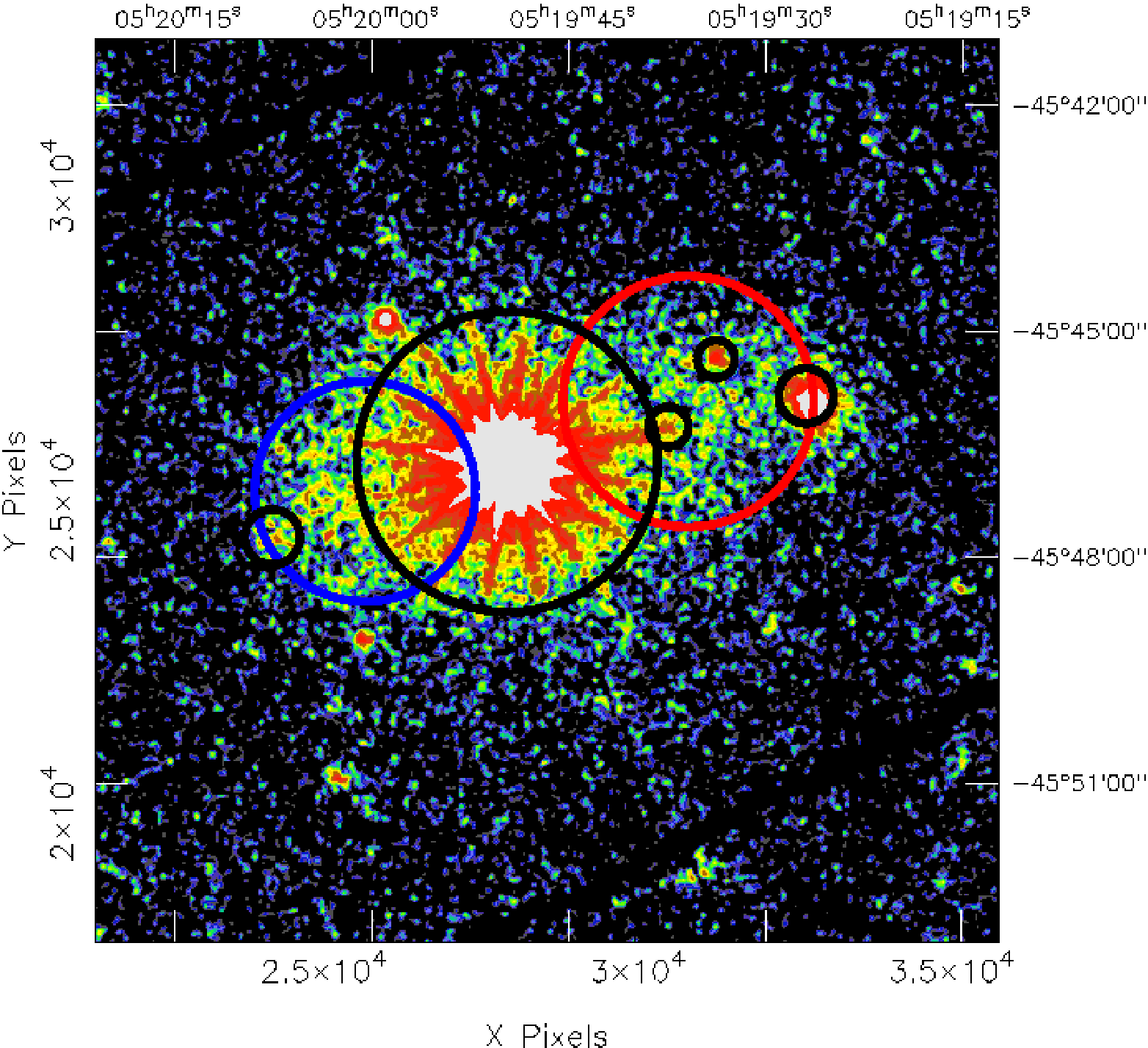}{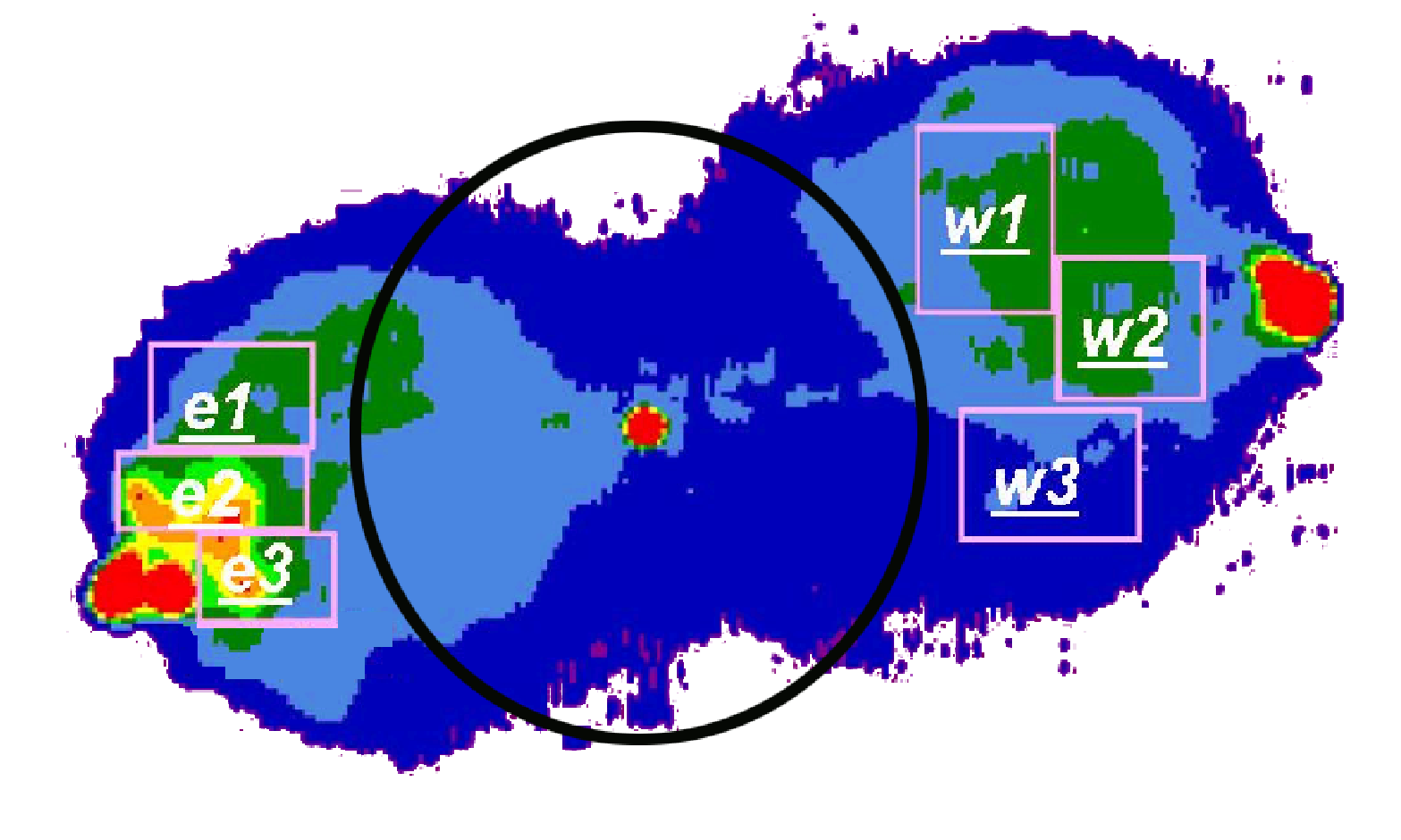}
\end{figure}

\begin{figure}[!hb]
\caption{ $B_{IC}$ ({\it Lower Panel}) and $k_e$ ({\it Upper Panel}) values of east and west sub-regions are plotted as a function of the radio flux ($F/V$) at 1.4 GHz, normalized by the relative volume.}
\centerline{\psfig{figure=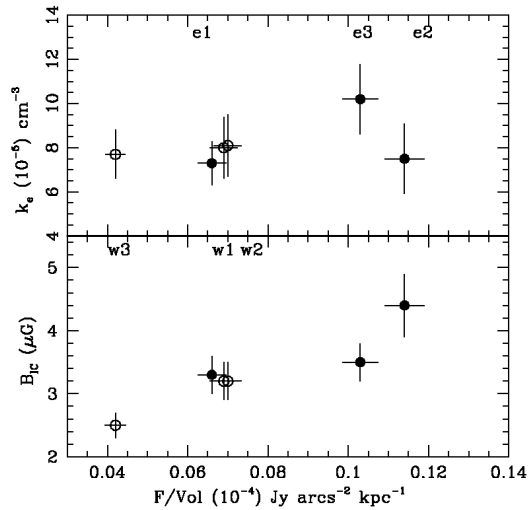,width=7cm}}
\label{fig_2}
\end{figure}

Combining the X-ray fluxes at 1 keV, $F_{1~keV}$, with the corrispondent radio fluxes, $F_{1.4~
GHz}$, derived at 20 cm from the VLA image produced by \citet{perl97}, the magnetic field $B_{IC}$ can be immediately estimated from the formula:
\begin{equation} 
B_{IC} = [\frac{F_{1.4~
    GHz}}{F_{1~keV}}\frac{C_{IC}(\alpha) (1+z)^{\alpha+3}}{C_{sin}(\alpha)}]^{\frac{1}{\alpha+1}}[\frac{\nu_{syn}}{\nu_{IC}}]^{\frac{\alpha}{\alpha+1}}.
\end{equation}   
where $C_{IC}$ and $C_{syn}$ are quantities depending only from $\alpha=\Gamma-1=\alpha_x=\alpha_R$, with $\alpha_x$ and $\alpha_R$  X-ray and radio spectral indeces, and  $V=A\times s$  is the volume, $A$ being the area (arcsec$^2$) projected on the sky plane  and $s$ the thickness of the analyzed region. From synchrotron radio and Compton scattered X-ray luminosities \citep[see][]{blum70}, also the other physical quantities, i.e. the electron density, $k_{e}$, magnetic field and particle energy densities, respectively $u_{m}$ and $u_{e+p}$, are calculated. The results are reported in Tab.~\ref{tab_2} together with the values found for the E and W regions, corresponding to the average physical conditions in the two lobes.\\
The two lobes (E and W regions) show very similar physical conditions, both for the magnetic field ($B_{IC}\sim3\mu G$) and the particle density $k_{e}$. In particular, from the ratio $u_{e+p}/u_{m}\sim50\pm 10$, it emerges that the energetic of the lobes appears to be dominated by the particles, hinting at a possible departure from the equipartition principle. Using the equipartition formula revised by \citet{bru97}, we calculated the equipartition magnetic fields, $B_{eq}$, for both the two lobes. We found that $B_{eq}$ overestimates $B_{IC}$ by a factor of $\sim3$.
The equipartition estimate depends on some poorly known physical parameters such as the minimum energy of the relativistic electrons and the proton to electron energy ratio. However in \citet{migl07} we have shown that our conclusion are not changed even by allowing these parameters to vary within a viable physical range.
Interesting results derive from the spatially resolved analysis in spite of quite large uncertainties. In fact, if no change in the electron density appears evident in the different subregions of the two lobes, on the other side, Tab.~\ref{tab_2} gives an indication that $B_{IC}$ increases behind the east hot spot (e2 region in Fig.~\ref{fig_1}) where the radio flux is higher. A variation of the magnetic field was also considered in the model proposed by \citet{hard05}. When the data from both lobes are combined together, the previous results are strengthen. Plotting $B_{IC}$ and $k_{e}$ as a function of the radio flux (divided for the corresponding sub-region volume), like in Fig.~\ref{fig_2}, we find a statistically significant variation of $B_{IC}$ ($\chi^{2}=4.7$ with a probability $p=3.0 \times 10^{-4}$). A {\it Spearman test} gives a correlation coefficient $r=0.8$ with a $s=0.0499$ significance: $B_{IC}$ seems to trace the variation of the radio flux density. Once again, this behaviour is not reflected in the $k_{e}$ trend (Fig~\ref{fig_2}, {\it Upper panel}), and the correlation test gives $r=0.31$ with $s=0.54$.\\
The possibility of variation of the spectrum of emitting electrons through the lobes is a critical issue as these variations may force $B_{IC}$~- radio trends.
Anyway, an analysis of the spectral index radio maps between 20 and 6 cm and between 6 and 2 cm (kindly provided by
R. Perley and shown in Fig. 5 of Perley et al. 1997) shows  negligible spectral index variations ($\Delta \alpha \leq0.05$) from a sub-region to another one, confirming the robustness of our result.\\
The resulting picture might indicate a dynamical decoupling between the magnetic field and the electron energy densities as found in 3C452 by Isobe et al. (2002, 2005). A possibility is that plasma and high energy electrons expand on different scales:  the first in the regions immediately beyond the hot spots, while the second on larger scales giving a fairly constant density distribution. This is in line with the picture, of a decreasing magnetic field with the increasing distance from the hot spots, proposed by \citet{blund00} to explain the discrepancy between the radiative lifetimes of electrons and dynamical ages of double
radio galaxies. The presence of a spatial gradient of the magnetic field was also suggested by \citet{wii90} to justify the lack of spectral aging in the case of the lobes of 3C 234.


\acknowledgements 
We are very grateful to R. Perley for kindly providing quantitatively analyzable
radio images of Pictor A. 


\begin{table}[!hb]
\caption{XMM-Newton fit to the Western, W, and Eastern, E, Lobe of Pictor A in the
  0.5-10 keV band. Two model are tested: a thermal emission from hot diffuse gas and a non-thermal radiation modeled with a pure power law. $N_{H}$ is fixed to the galactic value $N_{H}^{Gal}= 4.18\times10^{20}~cm^{-2}$. The results strongly suggest a non-thermal origin of the radiation \citep[see][]{migl07}.}
\label{tab_1}
\begin{center}
{\scriptsize
\begin{tabular}{l c c c c c c}
\tableline
              &\multicolumn{3}{c}{$Power Law$}                      & \multicolumn{3}{c}{$Thermal~Emission$}\\
              &                    &                    &                        &               &                   &\\
              &$\Gamma$            &$\chi^{2}(dof)$     &$F_{0.5-2~keV}$         &$kT$         &$\chi^{2}(dof)$    &$F_{0.5-2~keV}$\\
              &                    &                    &$(erg~cm^{-2}~s^{-1})$  &$(keV)$      &                   &$(erg~cm^{-2}~s^{-1})$\\
\tableline\\ 
W             &1.7$^{+0.2}_{-0.2}$ &29(38)              &$12\pm1\times10^{-14}$  &$6^{+3}_{-1}$  &39(38)             &$12\pm1\times10^{-14}$\\
              &                    &                    &                        &               &                   &\\
E             &1.8$^{+0.2}_{-0.2}$ &33(31)              &$9\pm1\times10^{-14}$   &$5^{+3}_{-2}$  &45(31)             &$8\pm1\times10^{-14}$\\
\noalign{\smallskip}
\tableline
\end{tabular}
}
\end{center}
\end{table}
\normalsize

\begin{table}[!ht]
\caption{Magnetic field and particle energy densities of the Western, W, and Eastern, E, Lobe and spatially resolved analysis results.}
\label{tab_2}
\begin{center}
{\scriptsize
\begin{tabular}{l c c c c c c}
\tableline
\noalign{\smallskip}
       &$F_{1.4GHz}$       &$F_{0.5-2~keV}$         &$Area$            &$B_{IC}$    &$k_{e}$            &$u_{e+p}/u_{m}$\\
       &$(Jy)$             &$(erg~cm^{-2}~s^{-1})$  &$(arcsec^{2})$    &$(\mu G)$   &$(\times10^{-5}~cm^{-3})$ &\\
\noalign{\smallskip}
\tableline
\noalign{\smallskip}
\multicolumn{7}{c}{\bf East lobe}\\
       &                 &                          &                  &             &                  &         \\
E      &$11.4\pm0.3$     &$9.0\pm1.0\times10^{-14}$ &$12776\pm383$     &$3.1\pm0.2$  &$8.2\pm1.0$       &$56\pm10$\\
       &                 &                          &                  &             &                  &         \\
e1     &$2.5\pm0.1$      &$1.9\pm0.3\times10^{-14}$ &$2992\pm90$       &$3.3\pm0.3$  &$7.3\pm1.0$       &$44\pm11$\\
       &                 &                          &                  &             &                  &         \\
e2     &$2.8\pm0.1$      &$1.2\pm0.2\times10^{-14}$ &$1824\pm55$       &$4.4\pm0.5$  &$7.5\pm1.6$       &$25\pm8$ \\
       &                 &                          &                  &             &                  &         \\
e3     &$2.5\pm0.1$      &$1.5\pm0.2\times10^{-14}$ &$1872\pm56$       &$3.5\pm0.3$  &$10.2\pm1.6$      &$55\pm12$\\
\tableline
\noalign{\smallskip}
\multicolumn{7}{c}{\bf West lobe}\\
       &                 &                          &                  &             &                  &         \\
W      &$13.7\pm0.4$     &$12.0\pm1.0\times10^{-14}$&$21408\pm642$     &$2.9\pm0.2$  &$6.5\pm0.8$       &$50\pm10$\\
       &                 &                          &                  &             &                  &         \\
w1     &$3.3\pm0.1$      &$2.4\pm0.3\times10^{-14}$ &$3640\pm109$      &$3.2\pm0.3$  &$8.0\pm1.4$       &$51\pm13$\\      
       &                 &                          &                  &             &                  &         \\
w2     &$2.8\pm0.1$      &$2.4\pm0.3\times10^{-14}$ &$3024\pm91$       &$3.2\pm0.3$  &$8.1\pm1.4$       &$52\pm13$\\
       &                 &                          &                  &             &                  &         \\
w3     &$1.5\pm0.1$      &$2.9\pm0.5\times10^{-14}$ &$3400\pm102$      &$2.5\pm0.2$  &$7.7\pm1.1$       &$81\pm17$\\    
\noalign{\smallskip}
\tableline
\end{tabular}
}
\end{center}
\end{table}
\normalsize

\begin{thebibliography}{}
\bibitem[Blumenthal \& Gould(1970)]{blum70}
Blumenthal, G. R., Gould, R. J., \newblock 1970, $RvMP$, 42, 237  
\bibitem[Blundell \& Rawlings(2000)]{blund00}
Blundell, K. M., \& Rawlings, S., \newblock 2000, AJ, 119, 1111
\bibitem [Croston et al.(2005)]{cro05}
Croston, J. H., et al., \newblock 2005, ApJ,626, 733 
\bibitem [Brunetti et al.(1997)]{bru97}
Brunetti, G., et al., \newblock 1997, $A\&A$, 325, 898
\bibitem [Grandi et al.(2003)]{gra03} 
Grandi, P., et al., \newblock 2003, ApJ, 586, 123
\bibitem[Hardcastle et al.(2002)]{hard02}
Hardcastle, M. J., et al., \newblock 2002, ApJ, 581, 948
\bibitem[Hardcastle \& Croston(2005)]{hard05} 
Hardcastle, M. J., Croston, J. H., \newblock 2005, MNRAS, 363, 649
\bibitem[Harris \& Grindlay(1979)]{harr79}
Harris, D. E., \& Grindlay, J. E., \newblock 1979, MNRAS 188, 25
\bibitem[Isobe et al.(2002)]{iso02}
Isobe, N., et al., \newblock 2002, ApJ, 580, L111
\bibitem[Isobe et al.(2005)]{iso05}
Isobe, N., et al., \newblock 2005, ApJ, 632, 781
\bibitem[Kraft et al.(2000)]{kra00}
Kraft, R. P., et al., \newblock 2000, ApJ, 531L, 9  
\bibitem[Migliori et al.(2006)]{migl06} 
Migliori, G., et al., \newblock 2006, ESASP 604, 645
\bibitem[Migliori et al.(2007)]{migl07}
Migliori, G., et al., \newblock 2007, ApJ accepted (arXiv:0704.2131)
\bibitem[Miley(1980)]{mil80}
Miley, G., \newblock 1980, ARA\&A 18, 165M
\bibitem[Perley et al.(1997)]{perl97} 
Perley, R. A., et al., \newblock 1997 $A\&A$ 328, 12
\bibitem[Reynolds et al.(2001)]{rey01}
Reynolds, C. S., Sebastian, H., Begelman, M. C., \newblock 2001, ApJ, 549L, 179
\bibitem[Tashiro et al.(2001)]{tash01}
Tashiro, M., et al., \newblock 2001, ApJ, 546, L19T
\bibitem[Wiita \& Gopal-Krishna(1990)]{wii90}
Wiita, J. P., \& Gopal-Krishna, \newblock 1990, ApJ, 353, 476
\end{thebibliography}
\end{document}